\documentclass[pre,aps,twocolumn,showpacs,superscriptaddress,floatfix]{revtex4-2}

\usepackage{mathtools}
\usepackage[usenames,dvipsnames]{xcolor}
\usepackage{amsmath}
\usepackage{amssymb,mathrsfs}
\usepackage{graphicx}
\usepackage[colorlinks=true]{hyperref}
\usepackage{soul}
\usepackage{xcolor}

\usepackage{float}

\newcommand{\prt}{\partial}
\newcommand{\al}{\alpha}

\newcommand{\ga}{\gamma}
\newcommand{\om}{\omega}

\newcommand{\mtc}{\mathcal}

\newcommand{\rav}{\right\rangle }
\newcommand{\lav}{\left\langle }
\newcommand{\sn}{\mathrm{sn}}

\begin{document}

\title{Undular bore theory for the modified Korteweg-de Vries-Burgers equation}

\author{L. F. Calazans de Brito}
\affiliation{Higher School of Economics, 20 Myasnitskaya ul., Moscow, 101000, Russia}

\author{A.~M.~Kamchatnov}
\affiliation{Higher School of Economics, 20 Myasnitskaya ul., Moscow, 101000, Russia}
\affiliation{Institute of Spectroscopy, Russian Academy of Sciences, Troitsk, Moscow, 108840, Russia}
\affiliation{Skolkovo Institute of Science and Technology, Skolkovo, Moscow, 143026, Russia}

\begin{abstract}
We consider nonlinear wave structures described by the modified Korteweg-de Vries equation
with taking into account a small Burgers viscosity for the case of step-like initial conditions. The
Whitham modulation equations are derived which include the small viscosity as a perturbation.
It is shown that for long enough time of evolution this small perturbation leads to
stabilization of cnoidal bores and their main characteristics are obtained. Applicability
conditions of this approach are discussed. Analytical theory is compared with numerical solutions
and good agreement is found.
\end{abstract}



\maketitle

\section{Introduction}

The modified Korteweg-de Vries (mKdV) equation
\begin{equation}\label{eq1}
  u_t-6\al u^2u_x+u_{xxx}=0
\end{equation}
appeared first in study of the famous KdV equation
\begin{equation}\label{eq2}
  u_t+6uu_x+u_{xxx}=0
\end{equation}
related with Eq.~(\ref{eq1}) by Miura transformation \cite{miura-68}. The existence of such a
transformation allowed the pioneers of the inverse scattering transform method to discover this
method \cite{ggkm-67,mgk-68} for the KdV equation, and it was extended later to many other
equations including mKdV equation \cite{wadati-72,wadati-73} (see also, e.g., books
\cite{nmpz-80,as-81,newell-85} and references therein). The mKdV equation is almost as widely
used in physical applications as the KdV equation. Actually, the Gardner equation
\begin{equation}\label{eq3}
  u_t+6\beta uu_x-6\al u^2u_x+u_{xxx}=0
\end{equation}
combining nonlinear properties of the KdV and mKdV equations can be transformed to
Eq.~(\ref{eq1}) by a simple change of variables. Besides that, in physical applications it
often happens that the coefficient $\beta$ is very small and can be neglected, so Eq.~(\ref{eq3})
reduces directly to the  equation. The Gardner equation and its simplified mKdV version
find applications to the theory of nonlinear waves in stratified fluids, for
example, for description of large amplitude internal waves \cite{gpt-97,hm-06,aosl-07}.

One of the most important and universal phenomena in nonlinear physics is formation and
evolution of dispersive shock waves (see, e.g., review articles \cite{eh-16,kamch-21} and
references therein). They are called {\it undular bores} in water wave physics and they were
observed in both surface and internal waves. Their theory was originated by Gurevich and
Pitaevskii \cite{gp-73} who represented such structures as modulated nonlinear periodic
waves which evolution is governed by the Whitham modulation equations
\cite{whitham-65,whitham-74}, and they gave two typical examples of solutions which
describe dispersive shock waves---evolution of an initial discontinuity and formation of a
shock after generic wave breaking for the KdV equation case. Whitham modulation equations
for the mKdV case were derived in Ref.~\cite{don-76}, however their application to the
theory of dispersive shock waves turned out to be a quite difficult task even in the case
of an initial discontinuity problem. The reason for this difficulty is that the mKdV equation
is not genuinely nonlinear \cite{kpt-08}, that is the modulus of the ``nonlinear velocity'' 
$6\al u^2$ has an
extremal (minimal) value at $u=0$ on the contrary to the KdV equation case where the
``nonlinear velocity'' $6u$ is everywhere a monotonous function of the wave amplitude $u$.
As a result, in KdV case an initial discontinuity can only evolve into two different
structures (rarefaction waves or cnoidal undular bores) whereas in mKdV case an initial
discontinuity evolves into eight different wave structures depending on the parameters of
the initial jump of $u$. Some particular results in this direction were obtained in
Ref.~\cite{marchant} and the full solution was given in Ref.~\cite{kklhgceg-12} in the
context of the Gardner equation (\ref{eq3}).

In Gurevich-Pitaevskii theory, dispersive shock waves are expanding with time wave
structures, so that in the initial discontinuity type problems the change of modulation
parameters per unit length decreases with time and can become at large enough time smaller
than some other physical parameters which were neglected in derivation of Eqs.~(\ref{eq1})
or (\ref{eq2}). For so large values of time, the neglected effects must be taken into
account in the modulation theory. For example, small dissipation stops infinite expansion
of undular bores and their length is stabilized at some value inverse proportional to the
viscosity coefficient in accordance with early ideas of Refs.~\cite{bl-54,sagdeev} about
the structure of undular bores in water waves physics and plasma. The corresponding modified
Whitham equations for the KdV theory with weak Burgers dissipation were derived in
Refs.~\cite{gp-87,akn-87} and they were applied in these papers to description of stationary
dispersive shocks whose characteristic length is defined by the small viscosity coefficient
$\ga$ in the KdV-Burgers equation
\begin{equation}\label{eq4}
  u_t+6uu_x+u_{xxx}=\ga u_{xx}.
\end{equation}
The extension of this theory on the mKdV-Burgers (mKdVB) equation
\begin{equation}\label{eq5}
  u_t-6\al u^2u_x+u_{xxx}=\ga u_{xx}
\end{equation}
was discussed qualitatively in Ref.~\cite{ehs-17}, however the modified Whitham equations
were not obtained for this case and the quantitative theory was not developed. The main
aim of this paper is to derive the Whitham modulation equations for the mKdVB case (\ref{eq5})
and to apply them to the theory of undular bores. To this end, we will use the direct
Whitham method \cite{whitham-65,don-76} developed further for perturbed KdV equation in
Ref.~\cite{kamch-16}. Its advantage is that it does not need development of quite involved
methods of the inverse scattering transform (see Ref.~\cite{kamch-04}).
We obtain analytical formulas
for the main characteristics of shock waves and confirm them by numerical solutions
of Eq.~(\ref{eq5}).

\section{Elementary wave structures in mKdVB equation theory}

Wave structures evolved from an initial discontinuity are typically combined from several
types of elementary wave structures and at first we shall consider them briefly.
For definiteness we shall confine ourselves to the case of positive coefficient $\al>0$
although a similar theory can be developed for the case of negative $\al$. Naturally,
the viscosity coefficient $\ga$ is positive.

\subsection{Rarefaction waves}

First we consider situations when a wave connects two trivial solutions $u=u_-$ on the left
and $u=u_+$ on the right from the initial discontinuity, and assume that during the evolution
the wave remains a smooth function
of $x$. Then we can neglect dispersive and dissipative effects proportional to higher order
derivatives of $x$ and describe such a wave in the simplest approximation with account of
only nonlinear effects proportional to the first order space derivative,
\begin{equation} \label{eq6}
            u_t - 6\al u^2 u_{x} = 0.
\end{equation}
The boundary conditions suggest that there are two characteristic functions, one for the
sound wave propagating along the plateau $u=u_-$, which has the characteristic
$x_l = -6\al u_-^2 t$, and the other for the sound wave propagating along the plateau $u_+$,
so that this edge moves according to the equation $x_r = -6\al u_+^2 t$. Consequently,
the solution consists of three parts: $u=u_-$ for $x<x_l$, $u=u_+$ for $x>x_r$, and
between these two regions we have an evident self-similar solution of Eq.~(\ref{eq6}),
\begin{equation} \label{eq7}
			u(x, t) =
			\begin{cases}
				u_{-}, \quad x<x_l, \\
                \pm \sqrt{\frac{x}{-6\al t}}, \quad  x_l < x < x_r,\\
				u_{+}, \quad  x> x_r.
			\end{cases}
\end{equation}
Obviously, such a solution exists only if the boundary values $u_{\pm}$ satisfy the
conditions $0 < u_+  < u_- $ or $0 > u_+ > u_-$. In both cases these rarefaction waves (RWs)
propagate to the left.

\subsection{Periodic solutions}

\begin{figure}[t]
	   \centering
	   \includegraphics[width=8cm]{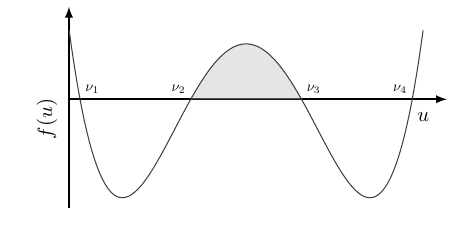}
	   \caption{Periodic solutions correspond to oscillations in the interval
$\nu_2\leq u\leq\nu_3$ where $f(u)\geq0$.}
	   \label{fig1}
\end{figure}

If the boundary values $u_{\pm}$ do not satisfy the above conditions, then the wave breaks
and undular bore forms. In Gurevich-Pitaevskii approach \cite{gp-73} they are represented
by modulated periodic solutions of Eq.~(\ref{eq5}), so at first we have to describe
the non-modulated solutions for zero dissipation.

We look for traveling wave solutions $u=u(\xi)$, $\xi = x - Vt$, of Eq.~\eqref{eq5}
with $\ga = 0$ and after two integrations we get
\begin{equation} \label{eq8}
		u_\xi^2 = \al u^4 + Vu^2 + 2Bu - 2A,
\end{equation}
where $A$ and $B$ are constants of integration. We assume that the polynomial in the right-hand side
has four real roots $\nu_i$, $i =1,2,3,4$, which are ordered according to inequalities
$\nu_1 \le \nu_2 \le \nu_3 \le \nu_4$, so Eq.~(\ref{eq8}) can be rewritten in the form
\begin{equation} \label{eq9}
			u_\xi^2 = \al (u-\nu_1)(u-\nu_2)(u-\nu_3)(u-\nu_4).
\end{equation}
The constants in these two equations are related by the expressions
\begin{align} \label{eq10}
	\color{red}
	&V = \al(\nu_1\nu_2 + \nu_1\nu_3 + \nu_1\nu_4 + \nu_2\nu_3 + \nu_2\nu_4 + \nu_3\nu_4), \notag \\
	&B = -\frac{\al}{2}(\nu_1\nu_2\nu_3 + \nu_1\nu_2\nu_4 + \nu_1\nu_3\nu_4 + \nu_2\nu_3\nu_4),  \\
	&A = -\frac{\al}{2}\nu_1\nu_2\nu_3\nu_4.  \notag
 \end{align}
and the roots $\nu_i$ are not independent of each other but connected by the formula
\begin{equation}\label{eq10a}
  \nu_1 + \nu_2 + \nu_3 + \nu_4 = 0.
\end{equation}
Periodic real solutions can only exist when $u$ oscillates between two consecutive roots
where the potential curve is positive, that is $\nu_2 \le u \le \nu_3$, as is shown in Fig.~\ref{fig1}.
Integration of Eq.~\eqref{eq9} with the initial condition $u=\nu_3$ at $\xi=\xi_0$ gives
\begin{equation} \label{eq11}
	   \xi - \xi_0 = \int_{u}^{\nu_3} \frac{du}{\sqrt{\al (u-\nu_1)(u-\nu_2)(u-\nu_3)(u-\nu_4)}}
\end{equation}
and standard calculation yields the expression
\begin{equation} \label{eq12}
	u = \frac{\nu_3(\nu_4 - \nu_2) - \nu_4(\nu_3 - \nu_2)\sn^2(\theta; m)}
{(\nu_4 - \nu_2) - (\nu_3 - \nu_2)\sn^2(\theta; m)},
\end{equation}
where $\sn(\theta, m)$ is the Jacobi elliptic sinus function,
\begin{equation}\label{eq12b}
  \theta = \frac{1}{2}\sqrt{\al (\nu_3 - \nu_1)(\nu_4 - \nu_2)}\xi,
\end{equation}
and
\begin{equation} \label{eq13}
	   m = \frac{(\nu_4 - \nu_1)(\nu_3 - \nu_2)}{(\nu_4 - \nu_2)(\nu_3 - \nu_1)}.
\end{equation}
Expression (\ref{eq12b}) allows us to define the wave number and the frequency of the periodic wave
in terms of parameters $\nu_i$:
\begin{equation}\label{eq13b}
  k=\sqrt{\al (\nu_3 - \nu_1)(\nu_4 - \nu_2)},\quad \om=kV,
\end{equation}
where $V$ is given by Eq.~(\ref{eq10}).

The cnoidal wave solution Eq.~\eqref{eq12} reduces to important particular solutions in special limits.
When $\nu_1 \rightarrow \nu_2$, so $m\to1$ and $\sn(\theta; m) \to \tanh\theta$, we arrive at the bright
soliton
\begin{equation} \label{eq14}
        u(\xi) = \nu_1 + \frac{\nu_3 - \nu_1}{\cosh^2\theta - \frac{\nu_3 - \nu_1}{\nu_4 - \nu_1}\sinh^2\theta}.
\end{equation}
propagating along a constant background $u = \nu_1$.

When $\nu_3 \rightarrow \nu_4$, we obtain the dark soliton
solution
\begin{equation}\label{eq15}
        u(\xi) = \nu_4 - \frac{\nu_4 - \nu_2}{\cosh^2\theta - \frac{\nu_4 - \nu_2}{\nu_4 - \nu_1}\sinh^2\theta}.
\end{equation}
propagating along a constant background $u = \nu_4$.

When $\nu_3 \rightarrow \nu_2$, we get $m \rightarrow 0$, so that the elliptical sinus becomes
the trigonometric one, $\sn(\theta; 0) = \sin\theta$, and we obtain a harmonic wave solution
oscillating with very small amplitude around $u=\nu_2$,
\begin{equation}\label{eq16}
        u(\xi) = \nu_2 + \frac{1}{2}(\nu_3 - \nu_2)\cos{(2\theta)}.
\end{equation}

At last, if we have simultaneously  $\nu_1 \rightarrow \nu_2$ and $\nu_3 \rightarrow \nu_4$,
it is convenient to change the initial condition in such a way that the integral (\ref{eq11})
takes the form
\begin{equation}\label{eq17}
	\xi = \int_{u}^{\nu_3} \frac{du}{\sqrt{\al}(u-\nu_2)(u-\nu_4)},
\end{equation}
and elementary integration yields
\begin{equation}\label{eq18}
		u(\xi) = \frac{1}{2}\{\nu_2 + \nu_4 \pm(\nu_2 - \nu_4)\tanh[\al (\nu_2 - \nu_4)\xi)]\}.
\end{equation}
It is important that due to Eq.~(\ref{eq10a}) the parameters are related by the
formula $\nu_2+\nu_4=0$ and therefore the left and right limiting values of $u$ have
opposite signs and their absolute values are equal to each other.
It is remarkable that exact solution of this type exists for the full Eq.~(\ref{eq5})
with account of dissipation~\cite{jmks-95} and we shall consider this modification
of the kink solution in the next Subsection.

\subsection{Kink}\label{kink}

Here we shall find the kink solution of Eq.~(\ref{eq5}) with $\ga\neq0$. As usual, we look for
a traveling wave solution $u=u(\xi)$, $\xi=x-Vt$, and assume that $u\to u_-$ as $\xi\to-\infty$.
Then trivial integration with account of our boundary condition gives
\begin{equation}\label{eq20}
  u_{\xi\xi}=\ga u_{\xi}+V(u-u_-)+2\al(u^3-u_-^3).
\end{equation}
Let we also have $u\to u_+$ as $\xi\to+\infty$, as it should be for a kink solution. Then
we get at once expression for the velocity
\begin{equation}\label{eq21}
  V=-2\al(u_-^2+u_-u_++u_+^2),
\end{equation}
and substitution of this expression into Eq.~(\ref{eq20}) gives
\begin{equation}\label{eq22}
  u_{\xi\xi}=\ga u_{\xi}+2\al(u-u_-)(u-u_+)(u+u_-+u_+).
\end{equation}
Now, following Ref.~\cite{jmks-95}, we assume that this equation has an integral in the form
$$
u_{\xi}=a(u-u_-)(u-u_+),
$$
that is
$$
u_{\xi\xi}=\frac{du_{\xi}}{du}\cdot\frac{du}{d\xi}=a^2(2u-u_--u_+)(u-u_-)(u-u_+).
$$
Substitution of these expressions into Eq.~(\ref{eq22}) yields
$$
a^2(2u-u_--u_+)=\ga a+2\al(u+u_-+u_+).
$$
Comparison of coefficients before $u$ gives $a^2=\al$ or
\begin{equation}\label{eq23}
  a=\pm\sqrt{\al}.
\end{equation}
Then the remaining terms give
\begin{equation}\label{eq24}
  u_-+u_+=\mp\frac{\ga}{3\sqrt{\al}}.
\end{equation}
At last, elementary integration of the equation
\begin{equation}\label{eq25}
  u_{\xi}=\pm\sqrt{\al}(u-u_-)(u-u_+)
\end{equation}
yields
\begin{equation}\label{eq26}
  u=\frac{u_-+u_+\exp[\mp\sqrt{\al}(u_+-u_-)(\xi-\xi_0)]}{1+\exp[\mp\sqrt{\al}(u_+-u_-)(\xi-\xi_0)]}.
\end{equation}
As one can see, the upper sign corresponds to the ``decreasing'' kink with $u_+<u_-$,
$u_++u_-=-\ga/(3\sqrt{\al})$ and the lower sign corresponds to the ``growing'' kink with $u_+>u_-$,
$u_++u_-=\ga/(3\sqrt{\al})$.

\section{Whitham modulation equations for mKdVB theory}

According to Whitham~\cite{whitham-65,whitham-74}, the modulation theory can be based on averaging
of the conservation laws for the equation under consideration over fast oscillations in the slightly
modulated cnoidal wave. The perturbed theory of the Whitham modulation method for the mKdVB equation
can be performed in the same way, as it was done for the KdVB equation \cite{kamch-16}.

Due to condition (\ref{eq10a}), in this theory there are three independent parameters which can be
chosen arbitrarily from the set $\nu_i$, $i=1,2,3,4$. Therefore we have to average three conservation
laws. However, it is convenient to replace one of them by the universal law of conservation of
`the number of waves' \cite{whitham-65,whitham-74}. Indeed, a slightly modulated wave can be considered
locally as a uniform one with the wave number and the frequency defined by the expressions
\begin{equation}\label{eq29}
  k=\theta_x,\qquad \om=-\theta_t.
\end{equation}
Consequently, they satisfy the conservation law
\begin{equation}\label{eq30}
  k_t+\om_x=0,
\end{equation}
where $k$ plays the role of `density of waves' and $\om$ is their `flux'. They are still expressed
in terms of local values of the modulation parameters $\nu_i$ by Eqs.~(\ref{eq13b}). Averaging
can be performed over a wavelength due to weakness of modulations,
\begin{equation}\label{eq31}
  \lav\phi\rav=\frac1L\int_0^L\phi dx=\frac{1}{L} \oint \frac{\phi(x,t)}{\sqrt{f(u)}} du,
\end{equation}
where $L=k^{-1}$ is the wavelength and $f(u)=u_x^2=\al\prod(u-\nu_i)$. Thus, the averaged
Eq.~(\ref{eq30}) can be written as
\begin{equation}\label{eq32}
  \lav k \rav _x + \lav \om \rav _t = 0,
\end{equation}
and it is easy to find two other conservation laws for the mKdVB case, so that in the averaged
form they read
\begin{align} \label{eq33}
    &\lav u \rav _t + \lav -2\al u^3 + u_{xx}\rav _x = \ga \lav R \rav, \notag \\
	&\lav u^2 \rav _t + \lav -3\al u^4 + 2uu_{xx} - u_x^2\rav _x = 2 \lav u R \rav,
\end{align}
where we denote by $R$ the general form of the perturbation term in the right-hand side
of the perturbed  mKdV equation. Of course, for Burgers friction we have $R=\ga u_{xx}$.

Following Refs.~\cite{whitham-65,whitham-74,don-76,kamch-16}, we express all averaged function
in terms of
\begin{equation} \label{eq34}
\begin{split}
	\mtc{W}(A,B,V)& = -\oint u_\xi du=-\oint \sqrt{f(u)} du\\
&=-\oint\sqrt{\al u^4+Vu^2+2Bu-2A}\,du,
\end{split}
\end{equation}
so that
\begin{equation}\label{eq35}
  \begin{split}
  &\mtc{W}_A=\oint\frac{du}{\sqrt{f(u)}}=\oint dx=L=k^{-1},\\
  &\mtc{W}_B=-\oint\frac{udu}{\sqrt{f(u)}},\\
  &\mtc{W}_V=-\frac12\oint\frac{u^2du}{\sqrt{f(u)}}.
  \end{split}
\end{equation}
Consequently, we get
\begin{equation}\label{eq36}
  \begin{split}
  & \lav u\rav=k\oint\frac{udu}{\sqrt{f(u)}}=-k\mtc{W}_B,\\
  & \lav \frac12u^2\rav=\frac{k}2\oint\frac{u^2du}{\sqrt{f(u)}}=-k\mtc{W}_V.
  \end{split}
\end{equation}
In view of the relation $u_{xx}=\frac12\frac{df}{du}$ we have $\lav u_{xx}\rav=0$. After simple
transformations with the use of the mKdV equation we can expressed all averaged quantities in
terms of the above expressions and arrive at
\begin{align} \label{eq37}
        &\left(-k \mtc{W}_B \right) _t + \left( -k V \mtc{W}_B + B \right) _x = \lav R\rav, \notag \\
	&\left( -k \mtc{W}_V \right)_t +\left( -k V \mtc{W}_V + A \right)_x =  \lav u R \rav, \\
	&\left( \mtc{W}_A \right)_t - V \left( \mtc{W}_A \right)_x = \mtc{W}_A V_x. \notag
\end{align}
These  equations can be rewritten in a more convenient way with the use of the differential operator
$\frac{D}{Dt} = \frac{\prt}{\prt t} + V \frac{\prt}{\prt x}$,
\begin{align}
        &\frac{D \mtc{W}_B}{Dt} = \mtc{W}_A\left(\frac{\prt B}{\prt x}-\lav R\rav\right), \notag \\
	&\frac{D \mtc{W}_V}{Dt} = \mtc{W}_A\left(\frac{\prt A}{\prt x} -  \lav u R \rav\right),\label{eq38}  \\
	&\frac{D \mtc{W}_A}{Dt} = \mtc{W}_A \frac{\prt V}{\prt x}. \notag	
\end{align}

As we mentioned in Introduction, the mKdV equation is not genuinely nonlinear. Therefore,
as in the case of the Gardner equation \cite{kklhgceg-12}, the relationship between physical parameters
$\nu_i$ and the most convenient modulation parameters used in the Whitham equations transformed to
the Riemann diagonal form is not single-valued. Correspondingly, we have to transform Whitham
equations (\ref{eq38}) for two different choices of independent variables. First, we choose
$\nu_1, \nu_2, \nu_3$ as such variables, so that $\nu_4$ is given by Eq.~(\ref{eq10a}) and
$d\nu_4 = -\left( d\nu_1 + d\nu_2 + d\nu_3 \right)$. Then differentials $dV, dA,$ and $dB$ of the
modulation parameters used in Eqs.~(\ref{eq38}) are equal to
\begin{equation}\label{eq39}
\begin{split}
  dV = &\al [ (\nu_4 - \nu_1 ) d\nu_1 + (\nu_4 - \nu_2) d\nu_2 \\
&+ (\nu_4 - \nu_3) d\nu_3   ] , \\
dB = &-\frac{\al}{2} [ \left( \nu_4 - \nu_1 \right) \left( \nu_2 + \nu_3 \right) d\nu_1 +
\left( \nu_4 - \nu_2 \right)  \left( \nu_1 + \nu_3 \right) d\nu_2 \\
&+  \left( \nu_4 - \nu_3 \right)  \left( \nu_1 + \nu_2 \right) d\nu_3  ],    \\
dA = &-\frac{\al}{2} [ \nu_2 \nu_3 \left( \nu_4 - \nu_1 \right) d\nu_1 +
\nu_1 \nu_3 \left( \nu_4 - \nu_2 \right) d\nu_2 \\
&+ \nu_1 \nu_2 \left( \nu_4 - \nu_3 \right) d\nu_3  ].
\end{split}
\end{equation}
Introducing the variables $w_i = \nu_4 - \nu_i$, we write  Eq~\eqref{eq38} in the form
\begin{equation}\label{eq40}
        \begin{split}
	 & \sum_{i=1}^{3} \mtc{W}_{A,\nu_i} \frac{D \nu_i}{Dt} = \al \mtc{W}_A  \left( w_1 \nu_{1,x} + w_2 \nu_{2,x} + w_3 \nu_{3,x} \right),  \\
	 & \sum_{i=1}^{3} \mtc{W}_{B,\nu_i} \frac{D \nu_i}{Dt} = -\frac{\al}{2} \mtc{W}_A  [ w_1(\nu_2+\nu_3) \nu_{1,x} \\
&+ w_2 (\nu_1+\nu_3)\nu_{2,x} + w_3(\nu_1+\nu_2) \nu_{3,x} ] -\mtc{W}_A  \lav R\rav,  \\
	 & \sum_{i=1}^{3} \mtc{W}_{V,\nu_i} \frac{D \nu_i}{Dt} = -\frac{ \al}{2} \mtc{W}_A ( \nu_2 \nu_3 w_1 \nu_{1,x} +  \nu_1 \nu_3 w_2 \nu_{2,x}\\
& +  \nu_1 \nu_2 w_3 \nu_{3,x}) -  \mtc{W}_A \lav uR \rav.
\end{split}
\end{equation}
To diagonalize the last system, we multiply the first, second and third lines by the constant
parameters $p, q,$ and $r$, correspondingly, sum the resulting equations, and choose $p,q,r$
in such a way, that the coefficient of
$\nu_{1,x}$ in the right-hand side vanishes and the coefficients of $\nu_{2,x}$
and $\nu_{3,x}$ are equal to each other. These conditions determine $p,q,r$ up to a
numerical factor and we take the following values:
        \begin{align}\label{eq41}
            &p = -(\nu_2 + \nu_3)(\nu_1 \nu_4 + \nu_2 \nu_3), \notag \\
            &q = -2(\nu_1 \nu_4 - \nu_2 \nu_3), \\
            &r = -4(\nu_2 + \nu_3). \notag
        \end{align}
After elementary transformations the resulting right-hand side of the sum takes the form
\begin{equation}\label{eq42}
\begin{split}
  &\mtc{W}_A\Big[\al (\nu_2-\nu_1)(\nu_3-\nu_1)(\nu_4-\nu_2)(\nu_4-\nu_3)\frac{\prt(\nu_2+\nu_3)}{\prt x}\\
  &+2(\nu_1\nu_4-\nu_2\nu_3)\lav R\rav+4(\nu_2+\nu_3)\lav uR\rav\Big]
  \end{split}
\end{equation}
Calculation of the coefficient before $D\nu_1/Dt$ gives
\begin{equation}\label{eq43}
  \begin{split}
  K_1&=p\mtc{W}_{A,\nu_1}+q\mtc{W}_{B,\nu_1}+r\mtc{W}_{V,\nu_1}\\
  &=-\frac{\nu_4-\nu_1}2\oint\frac{(p-qu-ru^2/2)du}{\sqrt{\al (u-\nu_1)^3(u-\nu_2)(u-\nu_3)(u-\nu_4)^3}}\\
  &=-(\nu_4-\nu_1)\oint\frac{d}{du}\sqrt{\frac{(u-\nu_2)(u-\nu_3)}{\al (u-\nu_1)(u-\nu_4)}}=0
  \end{split}
\end{equation}
Similar calculation of the coefficient before  $D\nu_2/Dt$ gives
\begin{equation}\label{eq44}
  \begin{split}
  K_2&=p\mtc{W}_{A,\nu_2}+q\mtc{W}_{B,\nu_2}+r\mtc{W}_{V,\nu_2}\\
  &=(\nu_4-\nu_2)(\nu_4-\nu_3)I_1,
  \end{split}
\end{equation}
where
\begin{equation}\label{eq45}
  I_1=\oint\sqrt{\frac{u-\nu_1}{\al (u-\nu_2)(u-\nu_3)(u-\nu_4)^3}}.
\end{equation}
As one can see, this expression is symmetrical with respect to interchange of $\nu_2$ and $\nu_3$,
so $K_3=p\mtc{W}_{A,\nu_3}+q\mtc{W}_{B,\nu_3}+r\mtc{W}_{V,\nu_3}=K_2$. Consequently, we
have obtained one of the modulation equations in the form
\begin{equation}\label{eq46}
  \begin{split}
  &(\nu_4-\nu_2)(\nu_4-\nu_3)I_1\left\{\frac{\prt(\nu_2+\nu_3)}{\prt t}+V\frac{\prt(\nu_2+\nu_3)}{\prt x}\right\}\\
  &=\mtc{W}_A\Big[\al (\nu_2-\nu_1)(\nu_3-\nu_1)(\nu_4-\nu_2)(\nu_4-\nu_3)\frac{\prt(\nu_2+\nu_3)}{\prt x}\\
  &+2(\nu_1\nu_4-\nu_2\nu_3)\lav R\rav+4(\nu_2+\nu_3)\lav uR\rav\Big],
  \end{split}
\end{equation}
and the other two equations can be obtained by cyclic permutations of $\nu_1,\nu_2{\color{red},}\nu_3$.

The terms, which do not depend on $R$, have diagonal form with respect to derivatives, so that three values of
any function of $\nu_1+\nu_2,\nu_1+\nu_3,\nu_2+\nu_3$ can serve as the Riemann invariants of the
resulting Whitham modulation equations. It is convenient to define them in the following way:
\begin{equation}\label{eq47}
  r_1=\frac14(\nu_2+\nu_3)^2,\quad r_2=\frac14(\nu_1+\nu_3)^2,\quad r_3=\frac14(\nu_1+\nu_2)^2
\end{equation}
and
\begin{equation}\label{eq48}
  \begin{split}
  &\nu_1=\sqrt{r_1}-\sqrt{r_2}-\sqrt{r_3},\quad \nu_2=-\sqrt{r_1}+\sqrt{r_2}-\sqrt{r_3},\\
  & \nu_3=-\sqrt{r_1}-\sqrt{r_2}+\sqrt{r_3},\quad \nu_4=\sqrt{r_1}+\sqrt{r_2}+\sqrt{r_3}.
  \end{split}
\end{equation}
The Riemann invariants $r_i$ are positive and we assume that they are ordered according to
inequalities $0<r_1\leq r_2\leq r_3$. Then the parameters $\nu_i$ are ordered as follows:
\begin{equation}\label{eq49}
  \nu_1\leq\nu_2\leq\nu_3<0<\nu_4.
\end{equation}
The phase velocity $V$ and elliptic modulus $m$ reduce to
\begin{equation}\label{eq50}
  V=-2\al(r_1+r_2+r_3),\quad m=\frac{r_3-r_2}{r_3-r_1},
\end{equation}
and the wavelength is given by the formula
\begin{equation}\label{eq51}
  L=\frac2{\sqrt{\al (r_3-r_1)}}K(m),
\end{equation}
$K(m)$ being the complete elliptic integral of the first kind. The integral (\ref{eq45}) can
also be expressed in terms of the Riemann invariants,
\begin{equation}\label{eq52}
  I_1=2 (\sqrt{r_2}-\sqrt{r_1})(\sqrt{r_3}-\sqrt{r_1})\frac{\prt L}{\prt r_1},
\end{equation}
and similar expressions can be obtained for its counterparts for equations derived from
Eq.~(\ref{eq46}) by cyclic permutations of $\nu_1,\nu_2{\color{red},} \nu_3$. As a result, we arrive
at the following form of the Whitham equations for the perturbed mKdV theory:
\begin{equation} \label{eq53}
		\frac{\prt r_i}{\prt t} + v_i \frac{\prt r_i}{\prt x} =
\frac{L}{\prt L/\prt r_i} \frac{\sqrt{r_1r_2r_3}\lav R\rav- r_i \lav uR \rav}{\prod_{j \neq i}(r_i - r_j)},
\end{equation}
where
\begin{equation}\label{eq54}
  v_i=\left(1-\frac{L}{\prt L/\prt r_i}\frac{\prt}{\prt_{r_i}}\right)V=V+\frac{2\al L}{\prt L/\prt r_i}
\end{equation}
are the standard Whitham velocities for the unperturbed mKdV equation \cite{don-76,ksk-04}.

Definitions (\ref{eq47}), (\ref{eq48}) of the Riemann invariants imply that in this case
a modulated wave oscillates in the region $\nu_2\leq u\leq \nu_3<0$ of its amplitude
(see Eq.~(\ref{eq49})). To get modulation equations for bores with positive values of
the amplitude, it is convenient to take $\nu_2,\nu_3,\nu_4$ as independent modulation
parameters, so that $\nu_1=-(\nu_2+\nu_3+\nu_4)$, and to define the Riemann invariants
by the formulas
\begin{equation}\label{eq55}
  r_1=\frac14(\nu_2+\nu_3)^2,\quad r_2=\frac14(\nu_2+\nu_4)^2,\quad r_3=\frac14(\nu_3+\nu_4)^2
\end{equation}
and
\begin{equation}\label{eq56}
  \begin{split}
  &\nu_1=-\sqrt{r_1}-\sqrt{r_2}-\sqrt{r_3},\quad \nu_2=\sqrt{r_1}+\sqrt{r_2}-\sqrt{r_3},\\
  & \nu_3=\sqrt{r_1}-\sqrt{r_2}+\sqrt{r_3},\quad \nu_4=-\sqrt{r_1}+\sqrt{r_2}+\sqrt{r_3}.
  \end{split}
\end{equation}
For $0<r_1\leq r_2\leq r_3$ the parameters $\nu_i$ are ordered according to
\begin{equation}\label{eq57}
  \nu_1<0<\nu_2\leq\nu_3\leq\nu_4
\end{equation}
and the variable $u$ takes positive values in the interval
\begin{equation}\label{eq58}
  0<\nu_2\leq u\leq\nu_3.
\end{equation}
The Whitham equations (\ref{eq53}) for this definition of the Riemann invariants remain the same.
Consequently, one solution of the Whitham modulation equations describes two different
modulated wave structures what is a characteristic feature of not-genuinely nonlinear wave
equations (other examples of such a behavior can be found in Refs.~\cite{kklhgceg-12,ik-17,ikcp-17}).

\section{Stationary bores in mKdVB theory}

As was mentioned in Introduction, after long enough time of evolution however small
dissipation stops expansion of undular bores and they acquire stationary profiles.
The corresponding theory for the KdV-Burgers equation was developed in
Refs.~\cite{gp-87,akn-87,kamch-16}. Here we shall obtain similar solutions for the case of
mKdVB theory following mainly to the method of Ref.~\cite{kamch-16}.

A stationary bore propagates with constant velocity $V$ without change of the profile
determined by the modulation variables $r_i=r_i(\xi)$, $\xi=x-Vt$. Such a stationary profile
is supported by the difference of the values of the wave variable $u$ at two infinities,
\begin{equation} \label{eq59}
			u(x, 0) \to
			\begin{cases}
				u_{-} , \quad \text{as}\quad x\to-\infty, \\
				u_{+} , \quad \text{as}\quad x\to+\infty. \\
			\end{cases}
\end{equation}
If there were no dispersion effects, we would get a jump-like viscous shock with velocity
determined by the Rankine-Hugoniot conditions (see, e.g., Ref.~\cite{whitham-74}).
Dispersion effects transform a jump-like transition between two levels of the $u$-variable
into an oscillatory bore, but the Rankine-Hugoniot conditions are still applicable \cite{ehs-17}.
Following Whitham's theory of weak shocks \cite{whitham-74}, we introduce the flux function
$Q=-2\al u^3$, so that the dispersionless limit of the mKdV equation takes the form the
conservation law
\begin{equation}\label{eq60}
  u_t+Q_x=0
\end{equation}
and then a shock wave propagates with velocity
\begin{equation}\label{eq61}
  V=\frac{Q(u_-)-Q(u_+)}{u_--u_+}=-2\al(u_-^2+u_-u_++u_+^2).
\end{equation}
(It is worth noticing that it coincides with velocity of kinks (\ref{eq21}) calculated with
account of viscosity what confirms the generality of the above argumentation). This velocity
must coincide with the constant velocity $V$ of the bore given by Eq.~(\ref{eq50}),
\begin{equation}\label{eq62}
  V=-2\al(r_1+r_2+r_3).
\end{equation}
Thus, in stationary solutions the sum of three Riemann invariants is constant and
Eqs.~(\ref{eq53}) reduce to
\begin{equation} \label{eq63}
        \frac{d r_i}{d \xi} = \frac{\sqrt{r_1r_2r_3}\lav R\rav- r_i \lav u R \rav}
        {2\al \prod_{i \neq j}(r_j - r_i)},\quad i=1,2,3.
\end{equation}
It is convenient to introduce symmetric functions of the Riemann invariants,
\begin{equation}\label{eq64}
		\sigma_1 = r_1 + r_2 + r_3, \quad \sigma_2 = r_1 r_2 + r_1 r_3 +r_2 r_3, \quad \sigma_3 = r_1 r_2 r_3.
\end{equation}
It is not hard to derive equations for them,
\begin{equation}\label{eq65}
  \frac{d \sigma_1}{d \xi} =0,\quad \frac{d \sigma_2}{d \xi} =\frac1{2\al}\lav uR\rav,\quad
  \frac{d \sigma_3}{d \xi} =\frac{\sqrt{\sigma_3}}{2\al}\lav R\rav.
\end{equation}
Consequently, $\sigma_1$ is an integral of motion, as it should be. The theory greatly
simplifies if $\lav R\rav=0$. In particular, it takes place for the Burgers viscosity:
$\lav u_{xx}\rav=(1/L)\left.(u_{xx})\right|_0^L=0$ due to periodicity of $u$ in the main
approximation. Then $\sigma_3=\mathrm{const}$ is also an integral of motion and we get
an ordinary differential equation for a sole dependent variable $\sigma_2$ or any
other variable changing along the bore. It is convenient to choose as such a variable
the modulus $m$. The Riemann invariants can be expressed as functions of $m$ in the
following way. The first and third equations (\ref{eq64}) give $r_1$ and $r_2$ as
functions of $r_3$:
\begin{equation}\label{eq66}
  \begin{split}
  r_1=\frac12\left[\sigma_1-r_3-\sqrt{(\sigma_1-r_3)^2-4\sigma_3/r_3}\right],\\
  r_2=\frac12\left[\sigma_1-r_3+\sqrt{(\sigma_1-r_3)^2-4\sigma_3/r_3}\right].
  \end{split}
\end{equation}
Then with the use of Eq.~(\ref{eq50}) for $m$ we find the formula
\begin{equation}\label{eq67}
  m=\frac{3r_3-\sigma_1-\sqrt{(\sigma_1-r_3)^2-4\sigma_3/r_3}}{3r_3-\sigma_1+\sqrt{(\sigma_1-r_3)^2-4\sigma_3/r_3}}
\end{equation}
which defines in implicit form the function $r_3=r_3(m)$, so that substitution of this
function into Eqs.~(\ref{eq66}) gives the functions $r_1=r_1(m),r_2=r_2(m)$. Differentiation
of $m$ by $\xi$ and substitution of Eqs.~(\ref{eq63}) with $\lav R\rav=0$ yield the
equation for $m$:
\begin{equation} \label{eq68}
        \frac{d m}{d \xi} =  -\Phi(m)
\end{equation}
Consequently, we obtain the solution in implicit form
\begin{equation}\label{eq69}
        \xi -\xi_0= \int_{m}^{1} \frac{dm}{\Phi(m)},
\end{equation}
where
\begin{equation}\label{eq70}
  \Phi(m)=\frac{ r_1(r_2 - r_3)^2 + r_2(r_1 - r_3)^2 + r_3(r_1 - r_2)^2}
  {2\al (r_1-r_2)(r_1-r_3)^3(r_2-r_3)} \lav uR \rav,
\end{equation}
and $\lav uR \rav$ can also be expressed in terms of the Riemann invariants, that is
as a function of $m$ ($\xi_0$ is the position of the soliton edge of the bore with $m=1$
at the initial moment of time).  This completes, in principle, solving
the Whitham equations for a stationary bore. When the function $m=m(\xi)$ is found,
it means that the dependence of the Riemann invariants $r_1,r_2,r_3$ on $\xi$ is also
known. Substitution of these functions into two sets (\ref{eq48}) and (\ref{eq56})
gives us two different dependencies of the parameters $\nu_i$, $i=1,2,3,4,$ on $\xi$.
This means that their substitution into the solution (\ref{eq12}) yields two
different modulated bores. The correct solution is distinguished by the boundary
conditions. Thus, now we are in position to classify all possible wave structures
supported by boundary conditions at infinities in the mKdV theory with account of small
Burgers viscosity.

\begin{figure}[t]
	   \centering
	   \includegraphics[width=8cm]{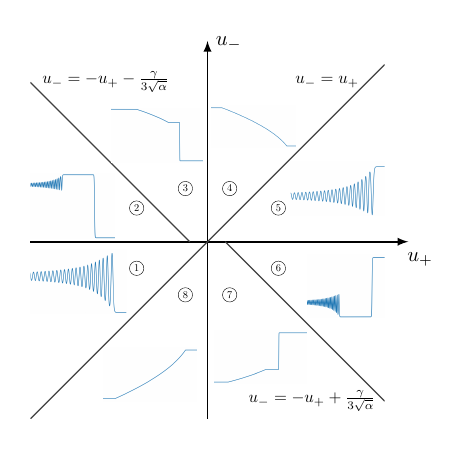}
	   \caption{Wave structures supported by the boundary condition $u_-$ as $x\to-\infty$
 and $u_+$ as $x\to+\infty$.}
	   \label{fig2}
\end{figure}

\section{Classification of wave structures for jump-like boundary conditions}

In the region of applicability of the Gurevich-Pitaevskii theory based on the Whitham
method of slow modulations of periodic solutions of the mKdV equation, the general diagram of
possible wave structures coincides qualitatively with the diagram obtained in
Ref.~\cite{kklhgceg-12} for the related Gardner equation without viscosity (see also
Ref.~\cite{ehs-17}). Taking viscosity into account leads to two modifications:
(i) undular bores become stationary and (ii) kinks' parameters are slightly changed as
it is shown in Section~\ref{kink}. The resulting diagram is shown in Fig.~\ref{fig2} and
here we shall derive analytical formulas for main characteristics of the wave structures
and compare them with numerical solutions of the mKdVB equation.

In the regions 1 and 5 in Fig.~\ref{fig2} we get just undular bores of different polarities.
Let us consider first the region~1 where $u_+<u_-<0$, so that $u$ oscillates in the negative
interval $\nu_2\leq u\leq \nu_3<0$. Correspondingly, we have to use formulas
Eqs.~(\ref{eq47}), (\ref{eq48}) relating $\nu_i$ and $r_j$. In the small amplitude limit
$x\to-\infty$ we have $\nu_2=\nu_3=u_-$ and $m\to0$, that is $r_2\to r_3$. Consequently,
we get at the left edge of the bore $r_1^-=u_-^2$, $r_2^-=r_3^-$, that is
\begin{equation}\label{eq71}
        \sigma_1  = u_- ^2 + 2r_2^-, \quad \sigma_3 = u_-^2(r_2^-)^2.
\end{equation}
At the soliton edge we have $m=1$, $r_2=r_1$, that is $\nu_1=\nu_2=-\sqrt{r_3}=u_+$,
that is $r_3^+=u_+^2$, so
\begin{equation}\label{eq72}
  \sigma_1  = 2r_2^++u_+^2, \quad \sigma_3 = (r_2^+)^2u_+^2.
\end{equation}
The values of these two constants of motion must be the same at both edges of the bore,
so simple calculations give the limiting expressions for the Riemann invariants at the
small amplitude edge,
\begin{equation}\label{eq73}
  r_1^-=u_-^2,\qquad r_2^-=r_3^-=\frac12u_+(u_++u_-),
\end{equation}
and at the soliton edge,
\begin{equation}\label{eq74}
  r_1^+=r_2^+=\frac12u_-(u_-+u_+),\qquad r_3^+=u_+^2.
\end{equation}
Naturally, their substitution into Eq.~(\ref{eq62}) reproduces the expression (\ref{eq61})
for the velocity of the bore. Besides that, we obtain the necessary expressions for the
constants of motion
\begin{equation}\label{eq75}
        \sigma_1 = u_-^2 + u_- u_+ + u_+^2, \quad \sigma_3 = \frac14u_-^2 u_+^2(u_- + u_+)^2.
\end{equation}

\begin{figure}[t]
	   \centering
	   \includegraphics[width=8cm]{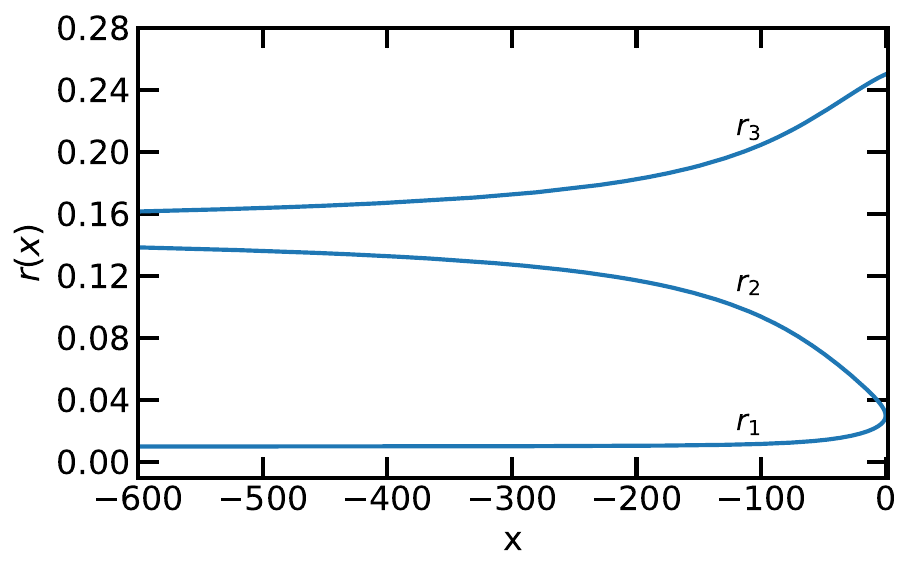}
	   \caption{Riemann invariants for the bores in regions 1 and 5 and the boundary conditions
$u_-=-0.1$, $u_+=-0.5$ in region 1 and $u_-=0.1$, $u_+=0.5$ in region 5.
The parameters of the equations are equal to
$\al=0.2$, $\ga=0.01$.
}
\label{fig3}
\end{figure}

\begin{figure}[t]
	   \centering
	   \includegraphics[width=8cm]{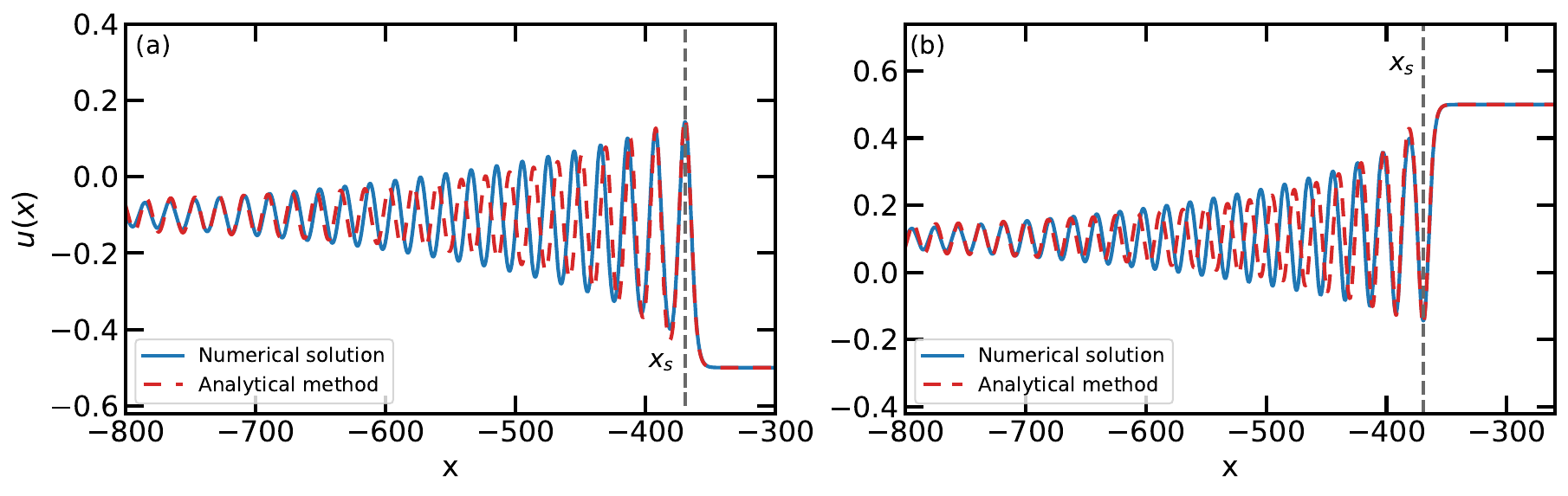}
	   \caption{The bore profiles for region 1 (a) and region 5 (b) found numerically
(solid blue lines) and analytically (dashed red lines).
In both cases the parameters of the mKdVB equation are equal to
$\al=0.2$, $\ga=0.01$ and the evolution time is $t=3000$. The boundary conditions are
$u_-=-0.1$, $u_+=-0.5$ in region 1 and $u_-=0.1$, $u_+=0.5$ in region 5.
}
\label{fig4}
\end{figure}

For averaging the Burgers friction term with $uR=\ga uu_{xx}$,
it is convenient to make a replacement $u \rightarrow 2v - s_1$, where
$s_1=\sqrt{r_1}+\sqrt{r_2}+\sqrt{r_3}$. The variable $v$ oscillates in the interval
$\sqrt{r_2} \leq v \leq \sqrt{r_3}$, so we obtain the expression
\begin{equation}\label{eq76}
\lav u u_{xx} \rav = - \frac{16}{L} \int_{\sqrt{r_2}}^{\sqrt{r_3}} \sqrt{Q(v)} dv
\end{equation}
where ${Q(v)} = \al (v - \sqrt{r_1}) (v - \sqrt{r_2})(v - \sqrt{r_3})(v - s_1)$.
The integral here can be expressed in term of the Jacobi elliptic integrals, but it is
convenient enough for practical calculations to keep it in this non-integrated form.

To find the criterium of applicability of our theory, we notice that it is correct as long
as the length $l$ of the whole bore is much greater than a typical local wavelength $L$
inside it. To estimate these two parameters, we turn to the small amplitude limit $\xi\to-\infty$
where the Riemann invariants are given by the formulas (\ref{eq73}). Then Eq.~(\ref{eq68})
reduces to
\begin{equation}\label{eq77}
  \frac{dm}{d\xi}= 4\ga m \quad \text{and} \quad m\propto
  \exp\left( 4\ga \xi\right),
\end{equation}
so the bore's length can be estimated as
\begin{equation}\label{eq78}
  l\sim\frac{1}{4\ga}.
\end{equation}
Substitution of Eqs.~(\ref{eq73}) into Eq.~(\ref{eq51}) gives according to the standard definition
$L=2\pi/k$ of the wavelength
\begin{equation}\label{eq79}
  L=\frac{\pi}{\sqrt{r_3-r_1}}=\frac{\sqrt{2}\pi}{\sqrt{(u_--u_+)|u_++2u_-|}}.
\end{equation}
Then the condition $L\ll l$ can be written in the form
\begin{equation}\label{eq80}
  u_--u_+\ll\frac{32\pi^2\ga^2}{|u_++2u_-|}.
\end{equation}
On the axis $u_-=0$ we get $-u_+\ll u_c=4\sqrt{2}\pi\ga$, and for $|u_-|\gg u_c$ we obtain
\begin{equation}\label{eq81}
  u_--u_+\ll\frac{u_c^2}{3|u_+|}\sim\frac{\ga^2}{|u_+|}.
\end{equation}
Thus, applicability region is separated from the line $u_+=u_-$ by a narrow strip formed by the
hyperbola boundary (\ref{eq80}).

In a similar way, in the region 5, where $u$ oscillates in the positive interval
$0<\nu_2\leq u\leq\nu_3$, we have to use the formulas (\ref{eq55}), (\ref{eq56}) relating
the Riemann invariants with the physical parameters of the wave. We obtain the same
formulas (\ref{eq73}) and (\ref{eq74}) for the limiting values of the Riemann invariants,
but for averaging the viscosity term we make a replacement $u=-2v+s_1$ and obtain
again the same formula (\ref{eq76}).

If we take symmetrical boundary conditions in regions 1 and 5 that differ only by signs,
then in both cases we get the same function $m=m(\xi)$ (see Eq.~(\ref{eq69})) and the
same plots of the Riemann invariants $r_1(\xi), r_2(\xi), r_3(\xi)$ shown in Fig.~\ref{fig3}.
Their substitution into Eqs.~(\ref{eq48}) or (\ref{eq56}) gives the dependencies $\nu_i=\nu_i(\xi)$,
$i=1,2,3,4$, for the modulation parameters of the bores in regions 1 and 5, correspondingly.
These functions $\nu_i=\nu_i(\xi)$ substituted into Eq.~(\ref{eq13}) yield the profiles of
bores in these two regions shown in Fig.~\ref{fig4} by red dashed lines. They are compared with
numerical solutions of the mKdVB equation and a quite good agreement is found, especially for
the positions and amplitudes of the leading solitons. The deviations of analytical plots from
numerical ones are caused by slow convergence of the wave structure to the stationary state.
Velocity of the shock is equal to Eq.~(\ref{eq61}) in the asymptotic state.

\begin{figure}[t]
	   \centering
	   \includegraphics[width=8cm]{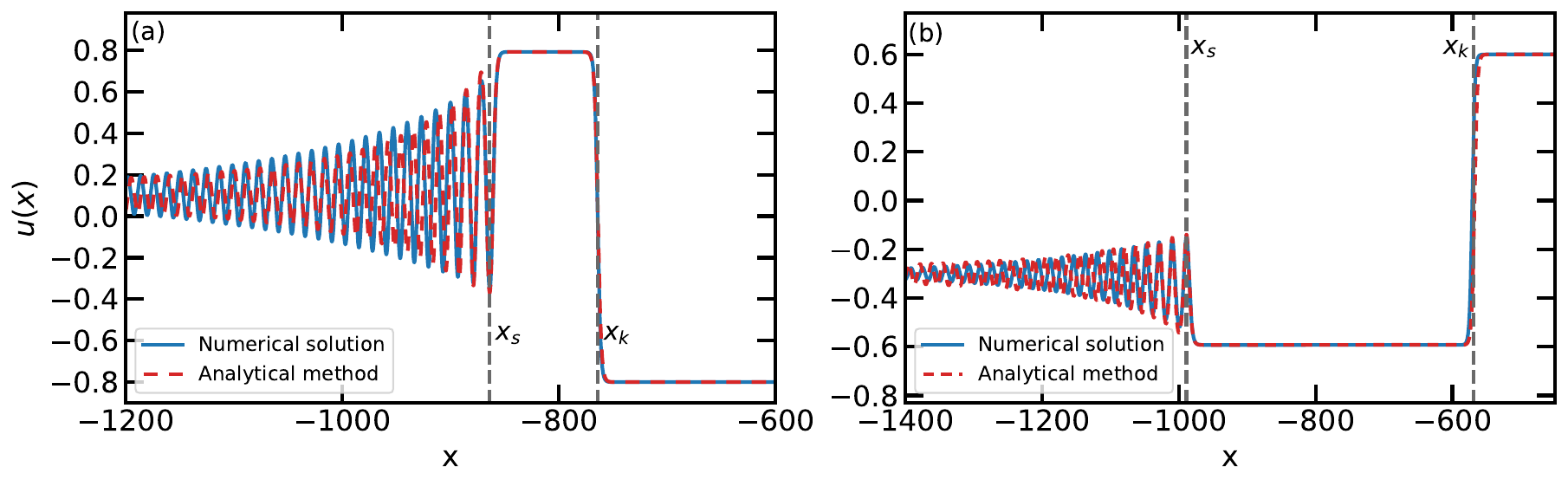}
	   \caption{The bore profiles for region 2 (a) and region 6 (b) found numerically
(solid blue lines) and analytically (dashed red lines).
In both cases the parameters of the mKdVB equation are equal to
$\al=0.2$, $\ga=0.01$ and the evolution times are $t=2500$ for region 2 and t = $4000$ for region 6. The boundary conditions are
$u_-=0.1$, $u_+=-0.8$ in region 2 and $u_-=-0.3$, $u_+=0.6$ in region 6.
}
\label{fig5}
\end{figure}

As was shown in Ref.~\cite{kklhgceg-12} for a similar Gardner equation, we cannot join the
boundaries $u_->0$ and $u_+<0$ by a single undular bore solution because the mKdV equation
is not genuinely nonlinear. In this case, the wave structure must contain a kink solution as
is shown in Fig.~\ref{fig2} for region~2 and for symmetrical region~6. In region~2 we have
a ``decreasing'' kink joining the right boundary $u_+<0$ with the intermediate plateau
\begin{equation}\label{eq82}
  u_*=-u_+-\frac{\ga}{3\sqrt{\al}}>u_-.
\end{equation}
This plateau is connected with the left boundary $u_-<u_*$ by the `negative' undular bore
which profile can be found in the same way as above with replacement $u_+\mapsto u_*$.
In particular, velocities of the kink and the bore are equal to
\begin{equation}\label{eq83}
  \begin{split}
  V_{kink}&=-2\al(u_*^2+u_*u_++u_+^2),\\
  V_{bore}&=-2\al(u_-^2+u_-u_*+u_*^2).
  \end{split}
\end{equation}
For separation of these two constituents in space, the difference
$$
V_{kink}-V_{bore}=2\al(u_--u_+)\left(u_--\frac{\ga}{3\sqrt{\al}}\right)
$$
must be positive. Hence, for realization of such a structure the left boundary must satisfy
the additional condition
\begin{equation}\label{eq84}
  u_->\frac{\ga}{3\sqrt{\al}}.
\end{equation}
If this condition is not fulfilled, then a combined rarefaction wave matched with a kink
is formed (see discussion of such situations in Ref.~\cite{ehs-17}.

In region~6 with $u_-<0$ and $u_+>0$ we get a structure with ``growing'' kink, so the
intermediate plateau has the amplitude
\begin{equation}\label{eq85}
  u_*=-u_++\frac{\ga}{3\sqrt{\al}}<u_-,
\end{equation}
and such a structure is realized for
\begin{equation}\label{eq86}
  u_-<-\frac{\ga}{3\sqrt{\al}}.
\end{equation}
We compared analytical and numerical solutions for regions~2 and 6 in Fig.~\ref{fig5}.
Again quite satisfactory agreement is observed.

\begin{figure}[t]
	   \centering
	   \includegraphics[width=8cm]{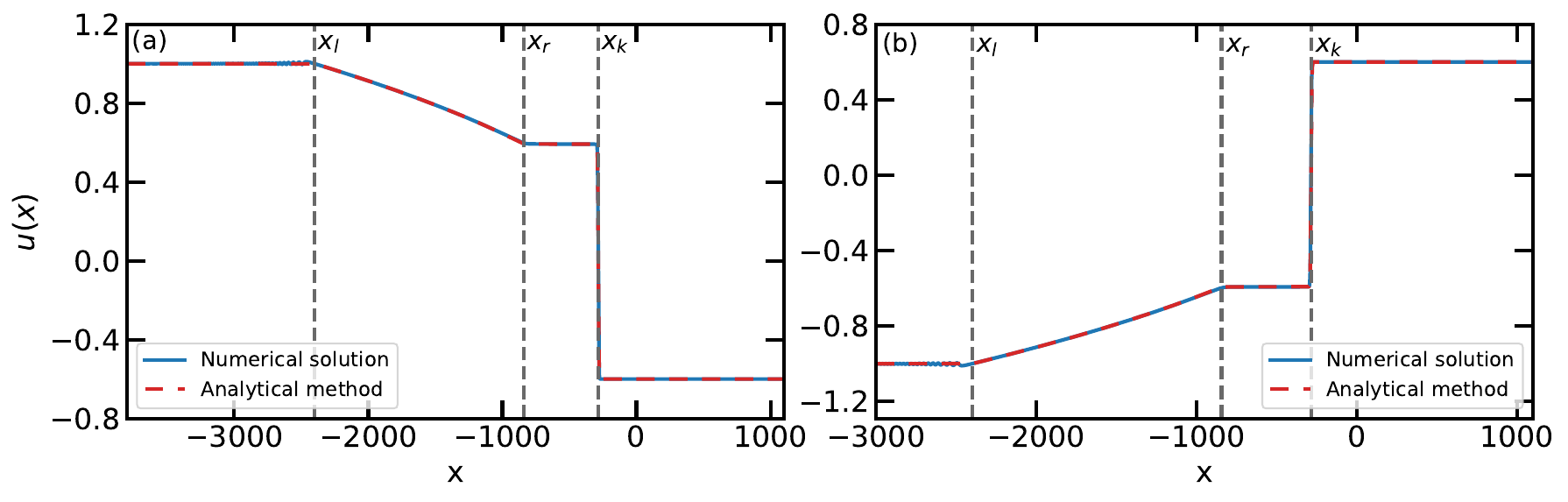}
	   \caption{The wave structures for regions 3 (a) and 7 (b) found numerically
(solid blue lines) and analytically (dashed red lines).
In both cases the parameters of the mKdVB equation are equal to
$\al=0.2$, $\ga=0.01$ and the evolution time is  $t=2000$. The boundary conditions are
 $u_-=1.0$, $u_+=-0.6$ in region 3 and $u_-=-1.0$, $u_+=0.6$ in region 7.
}
\label{fig6}
\end{figure}

It is clear that when $u_-$ reaches the level $u_-=u_*$, the cnoidal bore disappears and the
wave structure reduces to a sole kink. After further increase of $u_-$ we get into region~3
where the left boundary $u_-$ is joined with the plateau $u_*$ by a rarefaction wave (\ref{eq7}).
Its left edge propagated with velocity $V_{rw}^-=-6\al u_-^2$ and its right edge propagates with
velocity $V_{rw}^+=-6\al u_*^2$ which must be smaller than the kink's velocity. This gives the
condition
\begin{equation}\label{eq87}
  u_+<-\frac{2\ga}{3\sqrt{\al}} \quad  \text{or} \quad 0 > u_+ > -\frac{\ga}{6\sqrt{\al}}.
\end{equation}
for realization of such a structure in region~3. A similar structure in the symmetrical region~7
realizes for
\begin{equation}\label{eq88}
  u_+>\frac{2\ga}{3\sqrt{\al}} \quad  \text{or} \quad 0 < u_+ < \frac{\ga}{6\sqrt{\al}}.
\end{equation}
As one can see in Fig.~\ref{fig6}, the analytical theory agrees very well with the numerical
solutions for these two regions.

\begin{figure}[t]
	   \centering
	   \includegraphics[width=8cm]{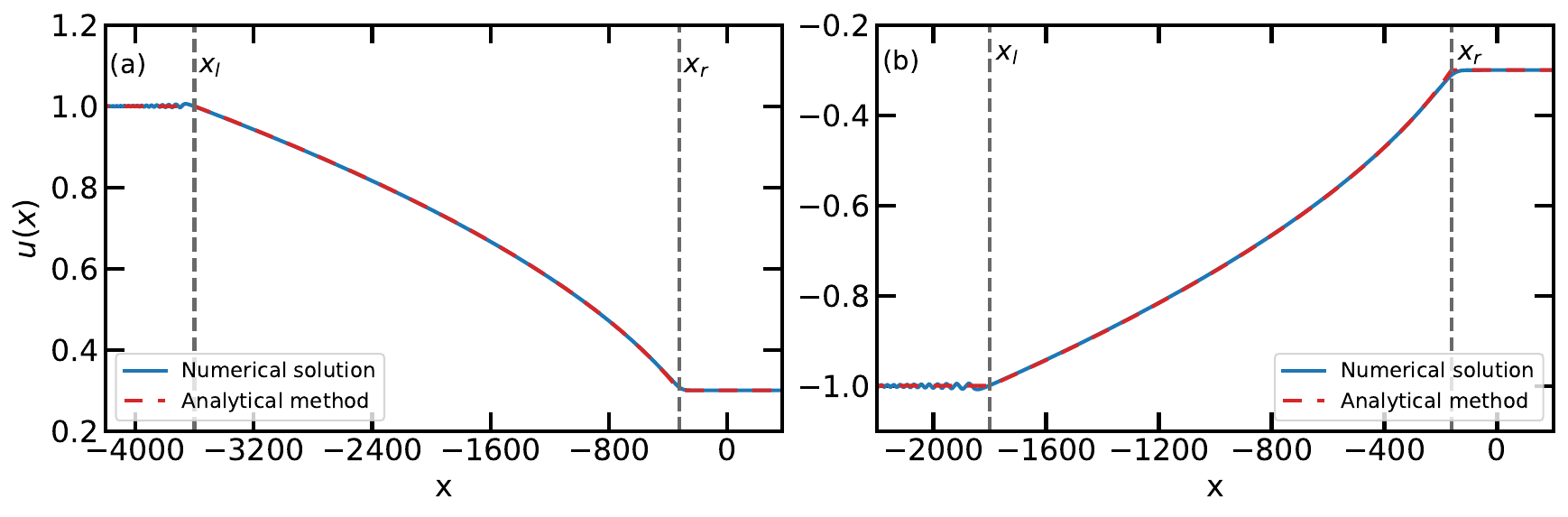}
	   \caption{ The wave structures for regions 4 (a) and 8 (b) found numerically
(solid blue lines) and analytically (dashed red lines).
In both cases the parameters of the mKdVB equation are equal to
$\al=0.2$, $\ga=0.01$ and the evolution   times are $t=3000$ for region 4 and $t=1500$ for region 8. The boundary conditions are
 $u_-=1.0$, $u_+=0.3$ in region 4 and $u_-=-1.0$, $u_+=-0.3$ in region 8.
}
\label{fig7}
\end{figure}

At last, in the regions~4 and 8 the boundary values $u_{\pm}$ have the same signs, so they are
connected by standard rarefaction waves with negligible influence of the Burgers friction
(see Fig.~\ref{fig7}). This completes the classification of possible wave structures supported by
different boundary conditions in the theory of the mKdVB equation.

\section{Conclusion}

The above theory confirms the general statement that weak dissipative effects stabilize
expanding evolution of dispersive shock waves, so after long enough time they converge to
stationary structures characterized by some finite length which is inverse proportional to the
viscosity coefficient. Appearance of the new parameter leads to some limitations on
applicability of the Whitham method used in the Gurevich-Pitaevskii approach to description
of bores. In particular, the condition that the size of the whole shock is much greater than
the typical wavelength inside the shock demands that the jump between the boundary conditions
is large enough. Since the mKdV equation is not genuinely nonlinear, we get combined wave
structures consisting of a kink and a cnoidal bore or a rarefaction wave. Small viscosity
leads to modification of the kink solution found in Ref.~\cite{jmks-95} and the condition
that the two structural elements of a combined structure propagate separately from each
other also leads to some limitations for boundary conditions. Although in case of small
viscosity these restrictions are not essential, one should keep in mind their existence
in practical application of the theory.

\begin{acknowledgments}

This research is funded by the research project FFUU-2021-0003 of the Institute of Spectroscopy
of the Russian Academy of Sciences (Sections~II, III) and by the RSF grant number~19-72-30028
(Section~IV, V).

\end{acknowledgments}

\end{document}